\documentclass[fleqn,usenatbib]{mnras}

\usepackage{newtxtext,newtxmath}

\usepackage[T1]{fontenc}

\DeclareRobustCommand{\VAN}[3]{#2}
\let\VANthebibliography\thebibliography
\def\thebibliography{\DeclareRobustCommand{\VAN}[3]{##3}\VANthebibliography}

\usepackage{graphicx}	
\usepackage{amsmath}	

\title[Measurement of dark matter subhalos mass]{An improved formalism for measuring the dark matter subhalos mass by its gravitational wake}

\author[M. E. Mosquera et al.]{
M. E. Mosquera,$^{1}$\thanks{E-mail: mmosquera@fcaglp.fcaglp.unlp.edu.ar}
K. J. Fushimi,$^{1}$
and M. J. de L. Dom\'{\i}nguez Romero$^{2}$
\\
$^{1}$Instituto de Astrof\'{\i}sica de La Plata (IALP, CONICET), Argentina\\
$^{2}$Instituto de Astronom\'{\i}a Te\'orica y Experimental, (IATE - UNC and CONICET CCT C\'ordoba), \\
Observatorio Astron\'omico de C\'ordoba, Universidad Nacional de C\'ordoba, Laprida 854, X5000BGR, C\'ordoba, Argentina.}

\date{Accepted XXX. Received YYY; in original form ZZZ}

\pubyear{2025}

\begin{document}
\label{firstpage}
\pagerange{\pageref{firstpage}--\pageref{lastpage}}
\maketitle

\begin{abstract}
We reexamined the framework used to determine the dark matter mass of a subhalo using the gravitational effects of its passage upon the stellar halo of a host. In particular, we aim to include different density distribution functions for the perturber and a non homogeneous background due to the host's halo gravitational potential. We have used a sample of K giant and RRLyrae stars based on Gaia DR3 data and two different set of simulations, to test the new formalism. From our analysis, we found that the inclusion of a nonhomogeneous background improves substantially the estimation of the subhalo dark matter mass. The methodology is not sufficiently sensitive to discriminate between different density distribution functions for the perturber, however, in the case of the observational data, the inclussion of the cloud’s circular velocity is a fundamental tool to complement the analysis. The results obtained including the host halo potential agree with the independent previous measurements or with the reported value for the subhalo dark matter mass in the simulations. These results show that it is possible to measure the mass of smaller subhalos through their gravitational wake, even for subhalos closer to their host, with distances lower than 30 kpc.
\end{abstract}




\section{Introduction}
\label{sec:intro}

Dark matter haloes provide the dominant gravitational potential wells where primordial inhomogeneities develop in today's galaxies. A vast corpus of observations in the local universe indicates that dark matter phase space distribution depends on dark matter particle properties. Given the hierarchical nature of the formation of dark matter haloes in the Cold Dark Matter (CDM) paradigm, an important number of substructures persist in it until they finally merge.

Those subhalos that retain gas and stars could be recognized as the satellite galaxies that populate the environment of main galaxies. The census of the Milky Way (MW), M31, and other nearby galaxies  provides a key constraint on dark matter properties and is still unfinished. The measurements of their physical properties are complicated by the strong environmental effects they suffer. Another salient prediction of the CDM model is an abundant population of dark subhalos that could only be detected by its gravitational effects.

One of the most crucial effects on the orbits and structure of the subhalos is the gravitational dragging caused by the dark matter distribution of its parent halo \citep{Weinberg:1986,Chandrasekhar:1943}. A main prediction for this work is the presence of a local wake. The local wake is an overdensity that follows the subhalo in its orbit, and its shape, intensity, and spatial distributions are dependent on dark matter properties \citep{Foote:2023}. This dark matter overdensity could also manifest as a stellar overdensity of the stellar halo and was predicted first in simulations for the Large Magellanic Cloud (LMC) by \citet{Garavito-Camargo:2019}. Subsequently, the Pisces Overdensity \citep{Sesar:2007,Watkins:2009} has been associated as the stellar counterpart to the LMC’s wake \citep{Belokurov:2019,Conroy:2021}. Notably due to the barycenter movement of the MW-LMC system, a global wake was measured in kinematics only \citep{Petersen:2021,Erkal:2021,Chandra:2025}, this only occurs for a very massive subhalo in relation to its host. \cite{Conroy:2021}, using K giants, was able to measure the stellar overdensity associated with the global wake, however \cite{Amarante:2024} did not find a statistical signicant signature of the global wake in the stellar density in their study with Blue Horizontal Branch stars.

As \citet{Buschmann:2018} noted, the observed stellar wake could be used to measure dark matter subhalo mass without any assumptions about its dynamical state, and potentially detect dark matter subhalos. These possibilities lead \citet{Fushimi:2024} to apply the likelihood formalism to measure the LMC and SMC system dark matter mass using updated Gaia DR3 information for a selected sample of stars with complete phase space information. This measurement is consistent with a range of mass estimations from different techniques. However, some simplifying assumptions of the Buschmann formalism remain untouched, namely i) a uniform background for the host (MW) dark matter potential, and ii) a simple Plummer spherical profile for the dark matter subhalo.

Therefore, this work explores the effects of accurately modeling the MW gravitational potential and different dark matter mass profiles for the subhalo. These improvements allow us to measure the mass and its errors more accurately and also open a pathway to measure the mass of smaller dark matter subhalos and possibly detect dark subhalos. This work is organized as follows. In Section \ref{formalismo} we develop the theoretical computation of an improved likelihood due to the presence of a wake generated by a dark matter subhalo moving in the stellar halo of its host galaxy. In Section \ref{data-results}, we introduce two examples of applying the new formalism: the well-studied case of the Magellanic Clouds wake and a similar merger extracted from the Auriga simulations dataset. Finally, in Section \ref{conclusiones}, we draw the conclusions and explore future developments.

\section{An improved formalism for the likelihood of the presence of a wake.}
\label{formalismo}

We have use the likelihood formalism to contrast the observational data with the theoretical model. In this case, the un-binned likelihood function is written as \citep{Buschmann:2018}
\begin{equation}
p(M_s,\theta)= e^{-N_\star(M_s)} \prod_{k=1}^{N_{d}} f\left(\bar{r}_k,\bar{v}_k\right)\,, \label{like}
\end{equation}
where $N_{d}$ is the number of star (data) in the region of interest (ROI), $N_\star$ is the theoretical amount of stars in the same region, $M_s$ is the subhalo dark matter mass of the perturber and $\theta$ stands for the extra constants involved in the computation. The function \mbox{$f\left(\bar{r}_k,\bar{v}_k\right)$} is the distribution function obtained by solving the collisionless Boltzmann equation, that depends on the position $\bar{r}$ and velocity $\bar{v}$ of each star. The coordinate system used to compute this function is centered in the perturber and its $x$-axis is aligned with the perturber's velocity (see \citet{Fushimi:2024} for details). The mass of the subhalo and its uncertainty were estimated from the logarithm of the likelihood function using a Markov Chain Monte Carlo (MCMC) method (emcee package; \citealt{emcee}), adopting the 16th, 50th, and 84th percentiles of the marginalised distributions to compute the uncertainties.


\subsection{Distribution function and $N_\star$ without a host halo potential}

We present the formalism needed to compute the likelihood function when the perturbed is immersed in an homogeneous background, with a constant and uniform density. The collisionless Boltzmann equation is 
\begin{equation}
\frac{\partial f}{\partial t} +\bar{v}\cdot \nabla_{\bar{r}} f-\nabla_{\bar{r}} \Phi_P \cdot \nabla_{\bar{v}} f=0 \, ,
\end{equation}
where $\Phi_P$ is the potential of the perturber. Following \citet{Buschmann:2018}, we propose 
\begin{equation}
f\left(\bar{r},\bar{v}\right)= f_0\left(\bar{v}\right) +f_1\left(\bar{r},\bar{v}\right)\, ,
\end{equation}
with 
\begin{equation}
f_0\left(\bar{v}\right)= n_0 \left(\sqrt{\pi} v_0\right)^{-3} e^{-\left(\bar{v}+\bar{v}_s\right)^2/v_0^2} \, ,
\end{equation}
where $\bar{v}_s$ is the dark matter subhalo velocity, \mbox{$v_0=\sqrt{2}\sigma_v$} with $\sigma_v$ the velocity dispersion and $n_0$ is the star density in the region of interest, of radius $R$. Therefore, at first order of the Newton's gravitational constant $G$, and for a time-independent potential, we can write
\begin{equation}
f\left(\bar{r},\bar{v}\right)= f_0\left(\bar{v}\right) -\frac{2 f_0\left(\bar{v}\right)}{v_0^2} \left. \int_0^\infty \frac{{\rm d} u}{u^3}\nabla_y \Phi_P(y)\cdot \left(\bar{v}+\bar{v}_s\right) \right|_{\bar{y}=\bar{r}-\frac{\bar{v}}{u}}  \, .
\end{equation}

The theoretical number of stars inside the ROI is defined as
\begin{equation}
N_\star = \int_{ROI} {\rm d}^3 r \, \int {\rm d}^3 v \,f\left(\bar{r},\bar{v}\right) \,. \label{ns_generica}
\end{equation}
The previous expression can be written as
\begin{equation}
N_\star= N_R +\Lambda F\left(\frac{v_s}{v_0}\right) \Bigg[\int_0^R {\rm d} x\,\rho_P(x)   \frac{x^2}{2}\left(3-\frac{x^2}{R^2}\right)  + \int_R^\infty {\rm d} x\,\rho_P(x)  x \, R \Bigg] \,,\label{ns}
\end{equation}
where $N_R= \frac{4}{3}\pi n_0 R^3$, $\Lambda=\frac{32 \pi^2}{3} \frac{R^2 n_0 G}{v_0 v_s}$, the function $F(a)$ is the Dawson function, and $\rho_P$ is the density of the perturber.


\subsection{Distribution function and $N_\star$ with a host halo potential}

If the perturber is a satellite of a host, the formalism described in the previous Section should be modified to take into account the effects of the host halo potential into the background. 

In this case, we compute the unperturbed distribution function $f_0$ from the Eddington equation \citep{Binney}, to take into account the non homogeneous background produced by the host's halo where the perturber is located
\begin{equation}
f_0\left(\epsilon\right)=\frac{1}{\sqrt{8}\pi^2} \left[\int_0^\epsilon \frac{{\rm d}^2 n_H}{{\rm d}\Psi^2} \frac{{\rm d}\Psi}{\sqrt{\epsilon-\Psi}} +\frac{1}{\sqrt{\epsilon}} \left(\frac{{\rm d} n_H}{{\rm d}\Psi}\right)_{\Psi=0}\right] \, .
\end{equation}
For the host halo considered as a NFW potential \citep{NFW97}, one has
\begin{align}
\Phi_H(r_H)&= \frac{-4\pi G  \, \rho_H \, R_H^3}{r_H } \ln \left(1+\frac{r_H }{R_H}\right)\, ,
\end{align}
where $R_H$ is the host scale factor, and $\rho_H$ is the host mass density. Following \citet{HunterPhDT}, we defined the adimensional quantities for the number density,
\begin{equation}
n_H(r_H)= n_0 \, \frac{R_H^3}{r_H \left(R_H+r_H \right)^2} \, , 
\end{equation}
and the potential as 
\begin{equation}
\tilde{n}_H(x)= \frac{n_H(x)}{n_0} \, , \hskip 1cm \tilde{\phi}_H(x)= \frac{\Phi_H(x)}{\epsilon_l} \, , 
\end{equation}
with \mbox{$x=r_H/R_H$} and \mbox{$\epsilon_l=4\pi G  \, \rho_H \, R_H^2$}. We can write \mbox{$\tilde{\Psi}(x)= -\tilde{\phi}_H(x)$} to find
\begin{align}
-\tilde{\Psi} e^{-\tilde{\Psi}}&=\left[-\tilde{\Psi} -\ln\left(1+x\right)\right] e^{-\tilde{\Psi} -\ln\left(1+x\right)} \, .
\end{align}
From this last expression one can obtain $x$ as a function of $\Psi$ using the W-Lambert function \citep{HunterPhDT}
\begin{equation}
x=-1-\frac{1}{\tilde{\Psi}} W_{-1}\left(-\tilde{\Psi} e^{-\tilde{\Psi}}\right) \, ,
\end{equation}
and therefore one can compute the distribution function as a function of \mbox{$\epsilon=\Psi(\bar{r}_H)-\frac{1}{2}v^2$}
\begin{equation}
f_0\left(\epsilon\right)=\frac{\epsilon_l^{-3/2}}{\sqrt{8}\pi^2} n_0 \int_0^{\epsilon/\epsilon_l} \frac{{\rm d}^2 \tilde{n}_H}{{\rm d}\tilde{\Psi}^2} \frac{{\rm d}\tilde{\Psi}}{\sqrt{\frac{\epsilon}{\epsilon_l}-\tilde{\Psi}}} \,. \label{f0}
\end{equation}

For the host halo considered as a Hernquist distribution, the potential is \citep{Hernquist90}
\begin{align}
\Phi_H(r_H)&= \frac{-G M_H}{R_H+r_H}\, ,
\end{align}
where $R_H$ is the host scale factor, and $M_H$ is the host mass. We defined the adimensional quantities as done previously to obtain \mbox{$x=r_H/R_H$} as a function of $\Psi$, resulting $x={\tilde{\Psi}}^{-1}-1$, and therefore one can compute the distribution function as done for the NFW potential, considering \mbox{$\epsilon_l= G \, M_H /R_H$}, using equation~(\ref{f0}).

\begin{figure}
\begin{centering}
\includegraphics[width=0.8\linewidth]{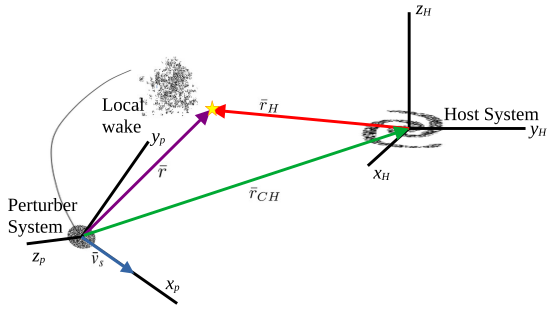}
\caption{Schematic relationship between the perturber ($p$) and the host ($H$) coordinate systems. The solid gray line represents the past orbit of the perturber. The star indicates the position of a generic star inside the ROI.} \label{sistemas}
\end{centering}
\end{figure}

The perturbed distribution function once again can be computed from the collisionless Boltzmann equation 
\begin{equation}
\frac{\partial f}{\partial t} +\bar{v}\cdot \nabla_{\bar{r}} f-\nabla_{\bar{r}} \Phi \cdot \nabla_{\bar{v}} f=0 \, , 
\end{equation}
where $\Phi$ is the total potential, that is the host dark matter Halo potential $\left(\Phi_H\right)$ plus the subhalo dark matter potential of the perturber $\left(\Phi_P\right)$. We propose \mbox{$f\left(\bar{r},\bar{v}\right)= f_0\left(\bar{r},\bar{v}\right) +f_1\left(\bar{r},\bar{v}\right)$}, and after some algebra considering a time-independent potential, the equation to solve can be written as 
\begin{equation}
f\left(\bar{r},\bar{v}\right)= f_0\left(\epsilon\right)+\left. \int_0^\infty \frac{{\rm d} u}{u^2}\nabla_{\bar{v}} f_0 \left(\epsilon_y\right) \cdot \nabla_{\bar{y}} \Phi(y) \right|_{\bar{y}=\bar{r}-\frac{\bar{v}}{u}} \, ,
\end{equation}
where \mbox{$\epsilon_y=\Psi_H\left(\bar{r}-\bar{r}_{CH}-\frac{\bar{v}}{u}\right)-\frac{1}{2}\left(\bar{v}+\bar{v}_s\right)^2$}, and $\bar{r}_{CH}$ is the position of the center of the host galaxy in the perturber frame \mbox{$(\bar{r}_H=\bar{r}-\bar{r}_{CH})$}. In Fig.~\ref{sistemas} we illustrate schematically the relationship between the perturber and host systems, along with a local wake and a generic star.

The theoretical number of star in the ROI, centered in the perturber, can be computed using Eq. \ref{ns_generica}, resulting
\begin{equation}
N_\star =\int_{ROI} {\rm d}^3 r \,n_H \left(\left|\bar{r}-\bar{r}_{CH}\right|\right) + \int_{ROI} {\rm d}^3 r \int {\rm d}^3 v f_1\left(\bar{r}, \, \bar{v}\right) \, .
\end{equation}


\subsection{Perturber dark matter densities}
In this work we have used four potentials to model the perturber dark matter subhalo, described in the following subsections.


\subsubsection{Plummer potential}
The density and potential are \citep{Plummer11}
\begin{equation}
\rho(r)=\frac{3 M_s}{4\pi R_s^3}\left(1+\frac{r^2}{R_s^2}\right)^{-5/2} \, , \hskip 1cm \Phi(r)=\frac{-G M_s}{\sqrt{R_s^2+r^2}} \, ,
\end{equation}
with \mbox{$R_s=1.62 \sqrt{M_s/10^8 M_{\sun }}$} \citep{erkal16}.


\subsubsection{NFW potential}

The density is \citep{NFW97}
\begin{equation}
\rho(r)=\frac{M_s}{4 \pi g_c }\frac{1}{r \left(R_s+r \right)^2}\, .
\end{equation}
The scale factor $R_s$ is a parameter fixed at 10 kpc, and the normalization constant is \citep{read:2019}
\begin{equation}
g_c=\ln \left(1+\frac{r_v}{R_s}\right)-\frac{r_v}{R_s+r_v}\,, \hskip 1cm
r_v= \left(\frac{3}{4\pi}\frac{M_s}{\Delta \, \rho_c}\right)^{1/3} \, , 
\end{equation}
where $M_s$ and $r_v$ are the virial mass and radius respectively considering \mbox{$\Delta=200$} and $\rho_c$ is the critical density of the Universe. 


\subsubsection{Hernquist potential}

In this case, one can write \citep{Hernquist90}
\begin{equation}
\rho(r)=\frac{M_s}{2 \pi}\frac{R_s}{r\left(R_s+r \right)^3}\, , \hskip 1cm 
\Phi(r)=\frac{-G M_s}{R_s+r} \, .
\end{equation}
We have taken the scale factor $R_s$ as a function of the mass, \mbox{$R_s=1.05 \sqrt{M_s/10^8 M_{\sun }}$} \citep{erkal16}. 


\subsubsection{Generic NFW potential}
In this case, the density can be written, as a function of the virial mass $M_s$, as \citep{walker09}
\begin{equation}
\rho(r)=\frac{M_s R_s^\beta}{4 \pi K_{norm}}\frac{1}{r^\gamma \left(R_s^\alpha+r^\alpha \right)^{\frac{\beta-\gamma}{\alpha}}}\, .
\end{equation}
The scale factor $R_s$ is a parameter fixed at 10 kpc, and the normalization constant is \citep{read:2019}
 \begin{align}
K_{norm}&=\frac{R_s^3}{3-\gamma} \left(\frac{r_v}{R_s}\right)^{3-\gamma} {}_2F_1\left(a, \, b, \, c; \, -\left(\frac{r_v}{R_s}\right)^\alpha\right) \,,  
\end{align}
where $a=\frac{3-\gamma}{\alpha}$, $b=\frac{\beta-\gamma}{\alpha}$ and $c=\frac{3+\alpha-\gamma}{\alpha}$. The parameter $\alpha$, $\beta$ and $\gamma$ can be related with the stellar and the dark matter mass according to \citep{li20,dicintio:2014}
\begin{align}
\alpha &= 2.94 - \log_{10} \left[\left(10^{X_{\star s}+2.33}\right)^{-1.08}+\left(10^{X_{\star s}+2.33}\right)^{2.29}\right]\,, \nonumber\\
\beta &= 4.23 +1.34 X_{\star s}+0.26 X_{\star s}^2 \,,  \\
\gamma &= -0.06 +\log_{10} \left[\left(10^{X_{\star s}+2.56}\right)^{-0.68}+10^{X_{\star s}+2.56}\right]\,, \nonumber
\end{align}
with $X_{\star s}=\log_{10}\left(M_\star/M_s\right)$.


\subsection{MW potential}

\begin{table}
\renewcommand{\arraystretch}{1.3}
\centering
\caption{NFW potential and density parameters for the MW as a host used in this work. $M_{MW}$ corresponds to the virial mass, $R_{MW}$ is the scale radius, $r^{MW}_{200}$ represents the virial radius and $c_{MW}$ is the concentration.} \label{modelo_MW}
\begin{tabular}{ccccc}
\hline 
Parameter & Model 1 & Model 2 & Model 3 & Model 4 \\ \hline
$M_{MW}\, [10^{12} M_{\sun }]$ &$1.29$ & $1.1$ & $1.1$ & $0.91$ \\
$R_{MW}$ [kpc] & $16.5$ & $16.5$ & $15.0$ & $15.62$\\
$r^{MW}_{200}$ [kpc] &$222.44$ & $210.93$ & $210.93$ & $198.18$ \\
$c_{MW}$ & $13.48$ & $12.78$ & $14.06$ & $12.68$ \\
\end{tabular}
\end{table}
To model the MW halo, we adopt, for the data and the Auriga simulations, a NFW density distribution function
\begin{eqnarray}
\rho_{MW}(r_H)&=&\frac{M_{MW}}{4 \pi g_{norm} }\frac{1}{r_H \left(R_{MW}+r_H \right)^2}\, ,
\end{eqnarray}
and considered four different cases, listed in Table~\ref{modelo_MW} \citep{vasiliev24,gala2}. Hereafter, Model 0 stands for the formalism without including the MW, that is with an homogeneous background. For the simulation of the LMC by \cite{Garavito-Camargo:2019}, we model the MW with a Hernquist profile, consistent with the simulation. The adopted parameters are described in Section \ref{gc-sim}.


\section{Applications of the formalism}
\label{data-results}

To study the dependence of dark matter subhalo mass on the different distribution functions proposed for DM subhalos, we used the gravitational response of the Milky Way’s halo stars to the passage of the Large Magellanic Cloud in its orbit. We performed this analysis using the data of Gaia Data Release 3 \citep{GAIA:DR3,Gaia:2016} and, afterward, in order to check the method employed we performed the same statistical test in two simulations data sets. 


\subsection{The observed wake generated by the Magellanic Clouds on the Milky Way halo Gaia stars.}

\begin{figure}
\begin{centering}
\includegraphics[width=\linewidth]{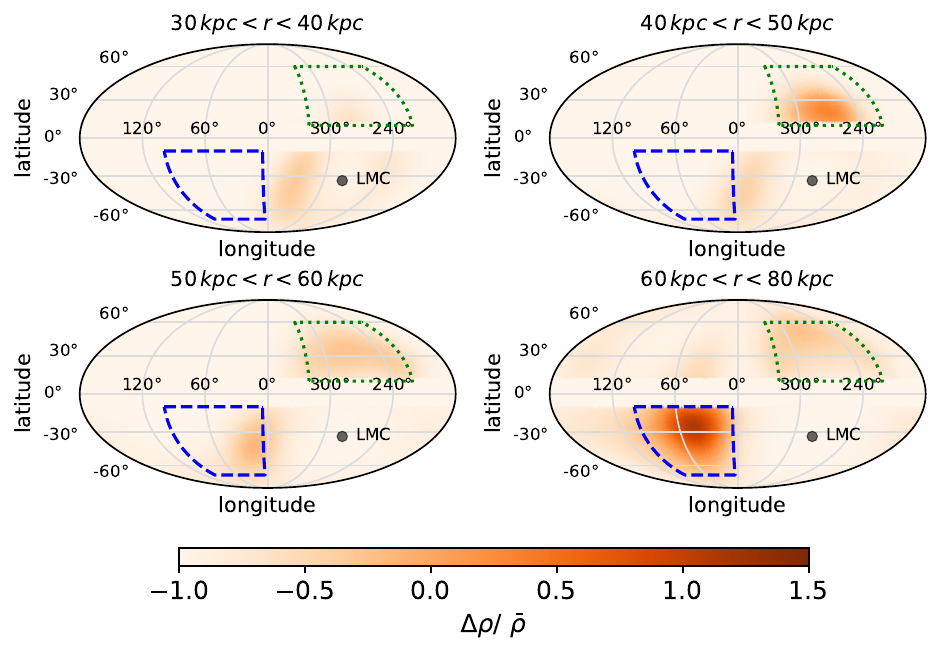}
\caption{Mollweide projection in galactic coordinates. The panels correspond to different distance from the CM of the MC. Green dotted line: region of the global wake; blue dashed line: region of the local wake.} \label{mollview}
\end{centering} 
\end{figure}

We used the reduced data of \cite{Fushimi:2024}. This data was obtained from the Gaia Data Release 3 \citep{GAIA:DR3,Gaia:2016} and reduced to created catalogs of K giants and RR Lyrae stars. For the data cleaning process, we accounted for dust extinction using an SFD map \citep{dustmap} and removed known objects, such as globular clusters and dwarf galaxies. To estimate distances, we used the MIST code to obtain isochrones \citep{MESA_1,MESA_2,MESA_3} for the K giants, which enabled us to calculate absolute magnitudes. For the RR Lyrae, we applied the metallicity absolute magnitude relation from \citet{Muraveva:2018} to obtain their absolute magnitudes. For stars without measured radial velocities, we estimated them using a RandomForestRegressor machine learning algorithm \citep{Breiman:2001,scikit} for the K giants, and a combined approach employing normalizing flows for data augmentation \citep{durkan2019neural,pzflow} and a RandomForestRegressor for prediction for the RR Lyrae. We focused on stars located between 30 and 100 kpc from the Galactic center, resulting in full phase-space catalogs of 6058 K giants and 2446 RR Lyrae stars. Figure \ref{mollview} presents the Mollweide projection of K giants and RR Lyrae stars in galactic coordinates. Each panel corresponds to a different range of distances to the center of mass (CM) of the Magellanic Clouds system (MC), inside the region of interest. We also highlight the regions of the local and global wakes. From the figure one can place the local wake between 60 to 80 kpc, and the global wake inside the range of 40 to 50 kpc both distances from the MC.

To obtain the LMC subhalo dark matter mass, we have performed a statistical analysis using the likelihood function obtained in the previous Section. For this test, we have considered as region of interest a spherical region centered in the CM of the MC system, of radius 100 kpc.

As described in \citet{Fushimi:2024}, the subhalo velocity was fixed at \mbox{$v_s=314.23 \, {\rm km}/{\rm s}$} \citep{vanderMarel:2002,Martinez:2019}. The star density was obtained from the observational data accordingly to the corresponding distribution function used in the analysis. The velocity dispersion was computed for each particular case. 

\begin{figure}
\begin{centering}
\includegraphics[width=\linewidth]{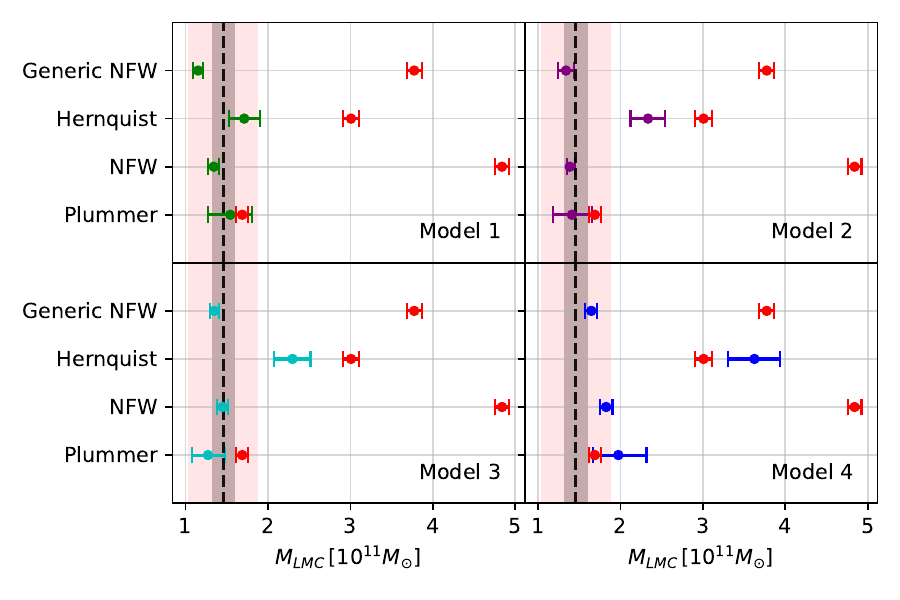}
\caption{Best-fit values for the LMC dark matter subhalo mass in units of $10^{11} M_{\sun }$. The colors represent the models used for the MW halo potential, red: no MW halo; green: Model 1; violet: Model 2; cyan: Model 3; blue: Model 4. 
The black-dashed line stands for the weighted mean value of the reported LMC dark matter subhalo mass in the literature, along with it $1\sigma$ and $3\sigma$ confidence levels.} \label{result-fig}
\end{centering}
\end{figure}

For the MW halo parameters we have considered four different cases, listed in Table~\ref{modelo_MW} \citep{vasiliev24,gala2}. The Model 0 stands for the case without the MW in the analysis, that is the background considered in this Model is uniform, this corresponds to the model used in \cite{Fushimi:2024}. In Fig.~\ref{result-fig} we show our results for the best-fit of the LMC dark matter subhalo mass, computed by using the MCMC method with a flat prior of $9<\log_{10} (M/M_{\sun })<11.8$, for each MW model and for each perturber potential considered. The mass for both NFW potentials are virial masses. 

As one can see from Fig.~\ref{result-fig}, our results considering a Plummer sphere as a perturber is in good agreement with the reported values extracted from the literature \citep{Correa:2021,koposov:2023,Shipp:2021,Vasiliev:2020,Erkal:2019,Penarrubia:2015}, independently of the presence of the MW halo potential. However, this changes when the perturber has a density function given by NFW or a generic NFW, where considering the MW potential turns fundamental to recover masses in good agreement with previous independent measurements. Also the general agreement for the subhalo mass between differently subhalos density profiles improves and is relatively independent of the exact model for the MW potential. In Table~\ref{resul} we present the values of the best fit of the dark matter subhalo mass of the LMC, used to plot Fig.~\ref{result-fig}. 

\begin{table*}
\renewcommand{\arraystretch}{1.3}
\centering
\caption{Best-fit values for the LMC dark matter subhalo mass, $M_{LMC}\, [10^{11} M_{\sun } ]$.} \label{resul}
\begin{tabular}{cccccc}
\hline
LMC potential &Model 0 & Model 1 & Model 2 & Model 3 & Model 4  \\ \hline
Plummer &  $1.69_{-0.07}^{+0.08}$ & $1.54 \pm 0.27$ & $1.41_{-0.23}^{+0.24}$ & $1.27_{-0.19}^{+0.20}$ & $1.97_{-0.31}^{+0.34}$ \\
NFW & $4.84_{-0.09}^{+0.08}$ & $1.34_{-0.07}^{+0.06}$ & $1.38_{-0.04}^{+0.07}$ & $1.45 \pm 0.07$ & $1.83 \pm 0.08$ \\
Hernquist & $3.01 \pm 0.10$ & $1.71 \pm 0.19$ & $2.33 \pm 0.21$ & $2.30 \pm 0.22$ & $3.63_{-0.32}^{+0.31}$\\
Generic NFW & $3.77 \pm 0.09$ & $1.15_{-0.07}^{+0.06}$ & $1.34_{-0.10}^{+0.09}$ & $1.35_{-0.06}^{+0.05}$ & $1.65_{-0.08}^{+0.07}$ \\ 
\end{tabular}
\end{table*}

To quantify the impact of the data uncertainties upon the best-fit mass, we performed random realizations obtained from Gaussian distributions (the means values correspond to the data and the standard deviations  are the uncentainties). For each realization, we fitted the subhalo dark matter mass and afterwards, we computed the mean and standard deviation over all realizations. The mean values agree with those in Table \ref{resul} within $1\sigma$, while the standard deviations are smaller than those from the MCMC method.

\begin{figure}
\begin{centering}
\includegraphics[width=\linewidth]{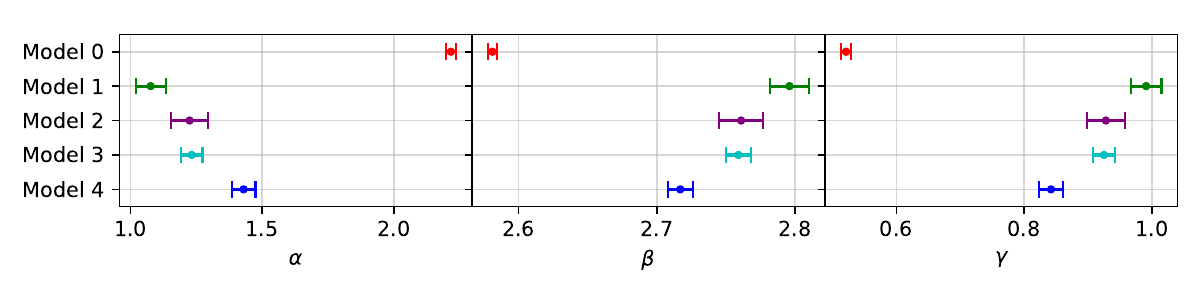}
\caption{Generic NFW parameters obtained form the best-fit values for the LMC dark matter subhalo mass. The colors represent the models used for the MW halo potential as in Fig. \ref{result-fig}. 
Left panel: $\alpha$; middle panel: $\beta$; right panel: $\gamma$.} \label{nfw-fig}
\end{centering}
\end{figure}

The generic NFW parameters were obtained form the best-fit values for the LMC dark matter subhalo mass for the four potential models of the MW considered and showed in Fig.~\ref{nfw-fig}. As one can see, the Model 1 favours a NFW potential, since $\alpha \approx 1$, $\beta \approx 3$ and $\gamma \approx 1$, for the perturber potential. For the others MW Models, the coefficients indicate that the perturber density is similar to a NFW. However, the model with uniform distribution density for the background (Model 0) has coefficients distant to any named distribution density. 

\begin{figure}
\begin{centering}
\includegraphics[width=0.8\linewidth]{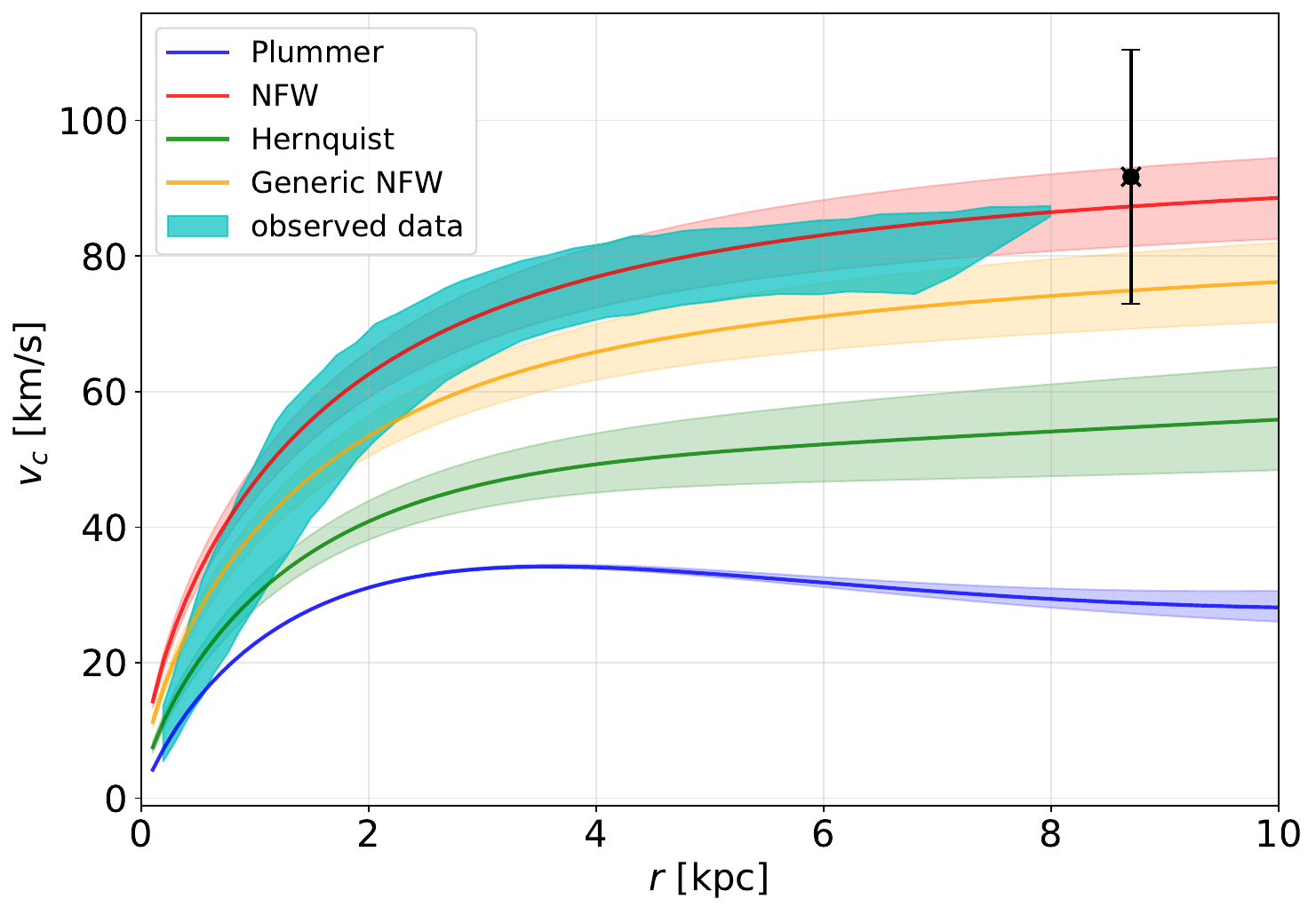}
\caption{Subhalo rotation curve for the best-fit masses obtained with Model 1. Cyan region: envelope of circular velocities obtained from literature; dot with the error bar: circular velocity of the LMC at $8.7$ kpc.}
\label{circular-vel}
\end{centering}
\end{figure}

Our method shows that including the MW potential improves the mass estimates; however, it is not sensitive enough to identify the density distribution that best fits the perturber. Therefore, we calculate the circular velocity using the standard expression $v_c(r) = \sqrt{G M(r)/r}$, where $M(r$) is the enclosed mass at radius $r$. We also include the contribution from the stellar disk of the LMC, modeled as an exponential disk \citep{Alves:2000}, to obtain the total circular velocity $v_c=\sqrt{{v_c}_{disk}^2+{v_c}_{halo}^2}$. Figure \ref{circular-vel} shows our results at the $3\sigma$ level for Model 1, where all calculated masses are in good agreement with the data in the literature. The cyan region corresponds to recent estimates of the LMC rotation curve derived from multiple tracers, including globular clusters, young main-sequence stars, red clump stars, and nearby field stars \citep{Niederhofer:2022,Dhanush:2024}, and represents a conservative observational constraint. We also include the value reported by \citet{vanderMarel:2014}. As shown, the NFW profile provides the best overall agreement with the full set of observational data.

\subsection{Analysis of the wake generated by the Magellanic Clouds on the Milky Way halo stars from simulations.}
\label{gc-sim}

For this analysis we used the data provided by \citet{Garavito-Camargo:2019}, that represent the LMC in an orbit arround the MW, where both the MW Halo, and the LMC subhalo were modeled as Hernquist profile. The dataset contains stars with a galactocentric distance larger than $50\, {\rm kpc}$. The LMC position and velocity in the galactocentric system was obtained from Table 8 (LMC 3) of \cite{Garavito-Camargo:2019} along with the data of \cite{Kallivayalil13} and \cite{vanderMarel:2014}. We consider the MW density profile as a Hernquist profile, whose parameters are \mbox{$M_{MW}=1.03 \times 10^{12} M_{\sun }$} (virial mass), \mbox{$R_{MW}=40.85\, {\rm kpc}$} (scale radius). For the subhalo, we consider different density profiles and for the Hernquist profile we have fixed the scale radius to 20 kpc as done in \cite{Garavito-Camargo:2019}.

\begin{figure}
\begin{centering}
\includegraphics[width=\linewidth]{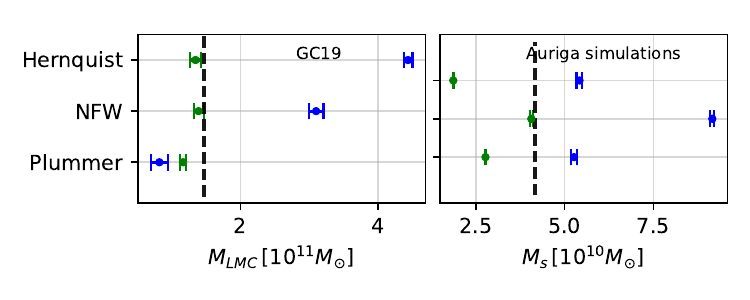}
\caption{ Best-fit values for the virial dark matter subhalo mass of the two data from simulations. Left panel: GC19 LMC and MW simulations; right panel: Most massive subhalo of the Halo 11 from Auriga simulations. The black-dashed line stand for the reported value of the virial dark matter subhalo mass, blue points: no Host Halo; green points: with Host Halo.} \label{graf_simulaciones}
\end{centering}
\end{figure}

To perform the likelihood analysis we have clean the data set by removing the LMC, excluding stars within a radius of $0.2\,r_{200}$ from its center. We selected randomly a smaller groups of 10000 stars inside the ROI (to have a sample of a similar size to the catalogue used in the previous section). We check our results with a different subsample of the data and the results were in agreement within $1\sigma$. The results for the virial dark matter mass of the subhalo of the LMC, obtained with the MCMC method with a flat prior of $10<\log_{10} (M/M_{\sun })<11.8$, are presented in the left panel of Fig.~\ref{graf_simulaciones}, for different profiles of the perturber. As one can see, the inclusion of the MW halo potential improves the estimation of the LMC virial dark matter subhalo mass, and the best fit for this case, $M_{LMC} = \left(1.35 \pm 0.08\right) \times 10^{11} M_{\sun }$, is consistent with the reported value if the perturber has a Hernquist density profile. 

\subsection{An example of an stellar wake in a Auriga halo that experienced a recent merger.}
\label{auriga-sim}

To test the presented formalism in more exigent conditions, that is a subhalo with a smaller dark matter mass than the LMC and closer to its host galaxy,  where the stellar density profile is higher, thus the signature of the wake should be stronger, we considered a subhalo from the Auriga simulations \citep{Grand:2024, Grand:2017}. We selected this subhalo since it experienced a recent merger with its host galaxy. 

In particular, for this work, we analized the Halo 11 and choose from it the satellite that has most massive dark matter subhalo (\mbox{$M_{s}=4.164 \times 10^{10} M_{\sun }$}), located at $23.69$ kpc from the center of the MW (both profiles are NFW). To determine the velocity of the subhalo, we identified the closest (bound) stars to this subhalo (stars within a radius of $r_{200}/5$ from the tabulated position of the subhalo, with $r_{200}$ its viral radius) and applied a 3-sigma clipping. The mean velocity of these closest stars can be taken as a good approximation for the subhalo velocity, resulting \mbox{$v_{s}=320.28$} km/s. We have removed not only the stars associated with the individual subhalos (within a radius of $r_{200}/3$ from each subhalo's center) but also the stars within a radius of $30$ kpc from the center of coordinates (center of the MW like host). Next, we performed the corresponding change of coordinate system to the perturber system, that is a roto-translation of both the coordinates and velocities of the stars (as describe in \citet{Fushimi:2024}). 

For the galaxy host, we have fit the circular velocity curve and consistent radial mass profile to find the NFW parameters of the host halo potential, resulting \mbox{$M_{MW}=1.10 \times 10^{12} M_{\sun }$} (virial mass) and \mbox{$R_{MW}=24.71\, {\rm kpc}$}.

To compute the likelihood analysis, we defined a region of interest with radius $1.5r_{200}$, centered in the subhalo, and considered the dark matter density of the subhalo as a NFW profile, with \mbox{$R_s=10$} kpc. In the right panel of Fig.~\ref{graf_simulaciones} we show our results, obtained with the MCMC method with a flat prior of $9<\log_{10} (M/M_{\sun })<11$, for different perturber's density profiles, along with the value of the dark matter subhalo mass reported by \citet{Grand:2024,Grand:2017}. Notably, including the host galaxy's dark matter halo profile significantly enhances the accuracy of the best fit of the dark matter subhalo mass. If one considered a NFW profile for the subhalo, the best fit is consistent with the reported one at $3\sigma$ level, $M_{s}^{with-MW}= \left( 4.06 \pm 0.05 \right) \times 10^{10} M_{\sun }$. This improvement is particularly predominant in the inner regions of the halo, where the density profile is higher. The enhanced accuracy in these regions is crucial for our understanding of the dynamics of subhalos near the host galaxy's center and the disk. 

\section{Conclusions}
\label{conclusiones}

In this work, we have addressed two salient shortcomings of the likelihood formalism for determining the subhalo mass that generates a local wake: accurately modeling the MW gravitational potential background and considering different dark matter mass profiles for the subhalo. The first consideration is important given the extended spatial nature of the phenomena, as is the case with the Large Magellanic Clouds where its local wake spans dozens of kpc. The former allows us to analyze smaller mass subhalos affected by gravitational tidal effects that could change their density distributions.

Our new analysis of the Magellanic Clouds' wake clearly shows a better agreement with previous independent mass determinations for the LMC dark matter subhalo mass. The study of different density profiles for the LMC for the nonhomogeneous background shows consistent masses for the subhalo (Fig.~\ref{result-fig}). Therefore, the methodology is not sensitive enough to discriminate between them. However, one can use the rotational curve to complement the analysis, as we shown in Fig.~\ref{circular-vel}, where the NFW profile is in agreement with the observational data. On the other hand, the improved formalism allows us to check that NFW density profile is consistently preferred for the MC's subhalo (Fig.~\ref{nfw-fig}). 

We also evaluated the MC's subhalo mass in a constrained simulation (Subsection \ref{gc-sim}), obtaining an improved measurement that agreed with the actual mass in the simulation. Additionally the new formalism introduced was tested in a less massive subhalo extracted from the Auriga simulations (Subsection \ref{auriga-sim}). This subhalo was near its MW-like host, and we obtained a good mass determination under more exigent circumstances (Fig.~\ref{graf_simulaciones}).

Finally, our improved techniques could lead to the detection or independent comprobation of any suspected dark subhalo in the Milky Way halo that possibly could only be detected by its gravitational effects, i.e. the local wake that generates in the halo stars distribution.

\section*{Acknowledgements}
This work was partially supported by grants from the National Research Council of Argentina. K. J. F. is a Post Doctoral fellow of the CONICET. M. E. M. and M. D. are members of the Scientific Research Career of the CONICET. 

This work has made use of data from the European Space Agency (ESA) mission {\it Gaia} (\url{https://www.cosmos.esa.int/gaia}), processed by the {\it Gaia} Data Processing and Analysis Consortium (DPAC) \url{https://www.cosmos.esa.int/web/gaia/dpac/consortium}), and code developed by the Gaia Project Scientist Support Team. Funding for the DPAC has been provided by national institutions, in particular, the institutions participating in the {\it Gaia} Multilateral Agreement. 

The authors want to thanks the referee for his/her coments that improved the manuscript. M.E.M. wants to thanks Francisco Azpilicueta for the helpful discussions. M.D. thanks Nicolás Garavito-Camargo by sharing the simulation data, and acknowledge the support of the CCA and the Flatiron Institute. We have used simulations from the Auriga Project public data release (\cite{Grand:2024,Grand:2017}) available at \url{https://wwwmpa.mpa-garching.mpg.de/auriga/data}.

This research has also made use of the following software: Astropy \citep{astropy}, Matplotlib \citep{matplotlib}, Pandas \citep{pandas}, Seaborn \citep{seaborn}, Healpy \citep{healpy}, SciPy \citep{scipy}, NumPy \citep{numpy}, The Jupyter Notebook \citep{jupyter-notebook}, gala \citep{gala,gala2}, Scikit-learn \citep{scikit}, cubepy \citep{cubepy}, pycuba \citep{pycuba}, emcee \citep{emcee}.

\section*{Data availability}
All Gaia and Auriga data used in this study are publicly available and can be download from \url{https://www.cosmos.esa.int/gaia} and \url{https://wwwmpa.mpa-garching.mpg.de/auriga/data} respectively.

\bibliographystyle{mnras}
\bibliography{bib}


\bsp	
\label{lastpage}

\end{document}